**Electronic Origin of the Inhomogeneous Pairing Interaction in the High-$T_c$ Superconductor Bi$_2$Sr$_2$CaCu$_2$O$_{8+\delta}$**

Abhay N. Pasupathy[1,*], Aakash Pushp[1,2,*], Kenjiro K. Gomes[1,2,*], Colin V. Parker[1], Jinsheng Wen[3], Zhijun Xu[3], Genda Gu[3], Shimpei Ono[4], Yoichi Ando[5], and Ali Yazdani[1,†]

[1]Joseph Henry Laboratories and Department of Physics, Princeton University, Princeton, New Jersey 08544, USA. [2]Department of Physics, University of Illinois at Urbana-Champaign, Urbana, IL 61801, USA. [3]Condensed Matter Physics and Materials Science, Brookhaven National Laboratory, Upton, NY 11973, USA. [4]Central Research Institute of Electric Power Industry, Komae, Tokyo 201-8511, Japan, [5]Institute of Scientific and Industrial Research, Osaka University, Ibaraki, Osaka 567-0047, Japan.

\* These authors contributed equally to this work

† To whom correspondence should be addressed email: yazdani@princeton.edu

Identifying the mechanism of superconductivity in the high-temperature cuprate superconductors is one of the major outstanding problems in physics. We report local measurements of the onset of superconducting pairing in the high–transition temperature ($T_c$) superconductor Bi$_2$Sr$_2$CaCu$_2$O$_{8+\delta}$ using a lattice-tracking spectroscopy technique with a scanning tunneling microscope. We can determine the temperature dependence of the pairing energy gaps, the electronic excitations in the absence of pairing, and the effect of the local coupling of electrons to bosonic excitations. Our measurements reveal that the strength of pairing is determined by the unusual electronic excitations of the normal state, suggesting that strong electron-electron interactions rather than low-energy (<0.1 volts) electron-boson interactions are responsible for superconductivity in the cuprates.

Central to the current debate on the mechanism underlying high-temperature superconductivity is the question of whether electron pairing in cuprates is caused by the exchange of bosonic excitations and can therefore be described by an extension of the Bardeen-Cooper-Schrieffer (BCS) theory, which has successfully explained phonon-mediated superconductivity in metals and alloys for the past 50 years (*1, 2*). Alternatively, it has been argued that the large Coulomb interaction in doped Mott insulators can result in a fundamentally different mechanism for pairing that cannot be approximated by a retarded boson-mediated interaction between electrons (*3-5*). In copper oxides, candidates for a BCS-like bosonic glue for pair binding include a magnetic resonance mode (near 40meV for hole-doped cuprates) (*6-8*), the spectrum of high-energy spin excitations (above 40 meV) (*9-12*), fluctuations around a quantum critical point (up to several hundred millielectron volts) (*13*), and phonons (e.g., those near 40 meV) (*14, 15*). Various spectroscopic measurements have probed these bosonic excitations and their coupling to electronic states in the cuprates (*14-20*); however, the connection between these excitations and the pairing mechanism has remained elusive (*5, 21*).

Particularly challenging is the fact that the pairing strength and the temperatures over which pairs form in cuprate samples such as $Bi_2Sr_2CaCu_2O_{8+\delta}$ are spatially inhomogeneous (*22-24*), a behavior that complicates the interpretation of macroscopically averaged experiments. Our technique allows for quantitative characterization of the electron-boson coupling and its correlation with the inhomogeneous pairing on the nanometer scale in the high-$T_c$ superconductor $Bi_2Sr_2CaCu_2O_{8+\delta}$. Our key finding is that while the coupling between electrons and bosons in the energy range of 20 to 120 meV influences the energy dependence of the pairing interaction, it does not control its local strength. We show instead that the local pairing strength is determined by asymmetric electron-hole excitations that are seen in the normal state properties well above the temperature where pairs first form.

**Lattice Tracking Tunneling Spectroscopy of Inhomogeneous Superconductors**

Electron-tunneling spectroscopy of superconductors is a powerful method for quantitative measurements of the onset of electron pairing and the boson exchange mechanism in superconductors (*25*). The pioneering measurements of MacMillan and Rowell (*26*) and their analysis (*27*) based on the extension of the BCS theory by Eliashberg (*28*), provided unequivocal evidence for a phonon-mediated mechanism of superconductivity in metals and alloys. In their study, the dependence of the tunnel conductance on the voltage was used to extract the phonon density of states and the strength of the electron-phonon coupling. These two quantities determine the transition temperature as well as the energy dependence of the pairing gap.

The success of tunneling as a quantitative spectroscopic probe in conventional superconductors relied on performing measurements in both the superconducting and normal state (*29*) in order to exclude the complication due to tunneling matrix elements as well as inelastic tunneling processes. For an inhomogeneous superconductor such as $Bi_2Sr_2CaCu_2O_{8+\delta}$, in which electronic states and the superconducting energy gap vary on the nanometer scale, similar experiments pose a technical challenge. It is necessary to track specific atomic locations on the lattice of the sample from low temperatures to temperatures above $T_c$ using a scanning tunneling microscope (STM) (*24, 30*). To enable such measurements, we have developed a thermally-compensated ultra-high vacuum STM system that can maintain thermal stability to better than 10 mK during spectroscopy experiments up to temperatures of 110 K.

The lattice tracking spectroscopy technique has been used to measure the evolution of tunneling conductance $dI/dV(r,V,T)$ with temperature in samples of $Bi_2Sr_2CaCu_2O_{8+\delta}$ that are overdoped ($T_c$=68 K, Fig. 1A and B) and optimally doped ($T_c$=93 K, Fig. 1D and E) at specific atomic sites. Spectra at different atomic sites show different energy gaps at low temperatures and evolve differently with increasing temperature (*24*). We find that all low temperature spectra in highly overdoped samples (Fig. 1A, B) evolve into spectra at high temperatures that are relatively featureless at low energies (<100 meV). In contrast, for optimally doped samples, we find that the spectra in most regions continue to have energy and temperature dependence at temperatures well above $T_c$ (Fig. 1 D and E).

When the tunneling conductance at high temperatures is weakly dependent on temperature and energy, following previous work on conventional superconductors (*29, 31*) we probe the effects of superconductivity by examining the ratio $R(r,V,T)$ between the tunneling conductance in the superconducting and normal states measured under the same STM setup conditions:

$$R(r,V,T) = \frac{[dI/dV(r,V,T)]_S}{[dI/dV(r,V,T)]_N} = \frac{N_S(r,V,T)}{N_N(r,V,T)} \qquad (1)$$

Here, $N_S(r,V,T)$ and $N_N(r,V)$ are the respective normal and superconducting density of electronic states at atomic site $r$ as a function of energy ($eV$). This ratio, which is independent of the tunneling matrix element, will be used to extract the temperature dependence of the energy gap and features associated with strong coupling of electrons to bosonic modes. Although our experimental technique can be extended to study samples at any doping, we focus on the quantitative analysis of the local temperature dependence of electronic states on overdoped samples. Photoemission (*17*) and Raman spectroscopy (*32*) studies on $Bi_2Sr_2CaCu_2O_{8+\delta}$ samples have shown the absence of pseudogap phenomena at similar doping levels, simplifying the analysis of the normalized spectra.

**Temperature dependence of the local pairing gap and quasiparticle lifetime**

The temperature evolution of the conductance ratio, *R(r,V,T)*, in overdoped $Bi_2Sr_2CaCu_2O_{8+\delta}$ samples at two representative locations of the sample with different low-temperature energy gaps are shown in Fig. 2A and B. Motivated by the fact that the low temperature ratio resembles that expected from a single energy gap in the spectrum, we compare these ratios with that expected from the thermally broadened density of states of a *d*-wave superconductor:

$$\frac{N_S(r,V,T)}{N_N(r,V)} = \frac{1}{\pi} \int dE \frac{df(E+V,T)}{dE} \int_0^\pi d\theta \, \text{Re} \frac{E - i\Gamma(r,T)}{\sqrt{(E - i\Gamma(r,T))^2 - \Delta(r,T)^2 \cos^2 2\theta}} \quad (2)$$

Here $\Delta(r,T)$ is the local *d*-wave gap amplitude (considered energy independent for low energies), $\Gamma(r,T)$ corresponds to the local inverse lifetime of the quasiparticle excitations (*33*), and *f(E,T)* is the Fermi function. Such an analysis neglects the complication due to the k-dependence of the bandstructure as well as higher order angular terms in the superconducting gap.

Given that the experimental spectra in our sample display a gap above $T_c$, one can question the appropriateness of using a single gap parameter to describe the normalized spectrum (Eq. 2). In general, the value of the energy gap can be determined by two different methods: from the slope of the normalized spectra near zero energy and from the energy at which the conductance spectra show a peak. If there were two different gaps - for example one dominating the nodal (near zero energy) and one dominating the antinodal (energy of the conductance peak) regions [as might be the case in underdoped $Bi_2Sr_2CaCu_2O_{8+\delta}$ samples (*17, 24, 32*)] - the two procedures would yield two different gap values. We find that the gap values obtained from the two methods agree at all locations on the sample, further justifying the use of a single gap (Eq. 2) to describe the spectra in our overdoped samples.

In general, we find that the model in Eq. 2 provides an excellent fit to the experimental data at low energies for all points on the overdoped samples (Fig. 2, A and B). Using this model, we can extract the local values of $\Delta(r,T)$ and $\Gamma(r,T)$, showing that at each point the gaps decrease monotonically with increasing temperature and close at a local temperature $T_p(r) > T_c$ (Fig. 2C). We find that $\Gamma(r,T)$ is much smaller than the gaps at all locations at low temperatures (Fig. 2D).

With increasing temperature, we find that the smaller gaps close first, with the largest gaps surviving to temperatures well above $T_c$. Regions of the sample with smaller gaps also show *R(r,V,T)* which exceeds that predicted from the local d-wave model in Eq.2 for $eV \approx \Delta(r,T)$. This behavior is likely due to localization effects experienced by the quasiparticles in small gap regions that cannot penetrate the larger gap regions (*34*). The extracted values of $\Gamma(r,T)$ are also in accord with this scenario – the regions with the smallest gaps show no lifetime broadening whereas the larger gap regions have a small lifetime broadening even at low temperatures (Fig. 2D). The variation in $\Gamma(r,T)$ is a consequence of the fact that the excitations in the large gap regions can decay into nearby regions with smaller energy gap, but not vice-versa.

As the smaller gaps begin to close with increasing temperature, we find that the $\Gamma(r,T)$ in the large gap regions begin to increase rapidly. Overall, the analysis of the experimental data for all regions on the sample demonstrates that the spatially averaged $\Gamma(T)$ shows a dramatic increase at a temperature $T \approx T_c$, when the sample loses long range phase coherence (Fig. 2D, inset). This observation is in accordance with previous macroscopically averaged measurements (*35*) on samples in the overdoped regime; however, previous measurements did not correlate inhomogeneous behavior of the gaps and quasiparticle lifetimes.

**Coupling of electrons to bosonic modes**

Having established that our measurements can be examined within the context of a local *d*-wave gap model, we turn to examining the deviation of the conductance ratio from this model for $E > \Delta$. Although other effects such as inelastic tunneling (*29, 36*) can cause such deviations, only strong coupling of electrons to bosonic modes is known to cause the superconducting state tunneling conductance to dip strongly below the normal state (*27, 29*). As illustrated in Fig. 3A, all points on the samples show a voltage range (around 50-80 meV) in which the conductance ratio *R(r,V,T)* is reduced below one and show systematic deviations from the local *d*-wave model. Analogous to previous work on conventional superconductors (*2*), these deviations provide a quantitative method to determine the strength of electron-boson coupling. The analysis of the spectra based on the features of *R(r,V,T)* instead of the bare *dI/dV(r,V,T)* or $dI^2/dV^2$*(r,V,T)* avoids complications due to the spatial variation of normal state features and tunneling matrix element variations (*21*). Although many previous studies, including those using an STM (*15, 37*), have examined electron-boson features in the 20-120 meV range, a quantitative comparison of the electron-boson coupling at different locations of the sample with different pairing gaps has not been accomplished. The comparison of *R(r,V,T)* at different locations allows us to quantitatively evaluate the role of bosonic features in the development of the pairing gaps and their inhomogeneity.

To study the relative strength of electron-boson coupling at different locations on the sample, we consider that the strong coupling to a bosonic mode at energy $\Omega$ in a superconductor results in features in the conductance ratio at $eV = \Delta + \Omega$ (*15, 29, 38*). As the pairing gap is locally varying, we plot the *R(r,V,T)* as a function of $eV - \Delta(r)$ for different atomic sites on the sample with low temperature $\Delta(r)$ ranging between 15 and 32 meV in Fig. 3B. This figure demonstrates that different locations on the sample show similar *R(r,V,T)* curves in magnitude and shape, once we take into account their varying pairing gaps. The only significant difference between the spectra occurs at low energies, where lifetime broadening effects as well the angle dependence of the electron-boson coupling, play a role. A measure of the energy $\Omega$ of the bosonic mode is the energy at which the dip occurs in the spectrum. We find that the average energy of this dip is 35 ± 3 mV. A quantitative measure of the strength of the local coupling constant is the root mean square (RMS) deviation of *R(r,V,T)* from the weak-coupling *d*-wave model (Eq.2) in the energy range 20 to 120 meV beyond the gap. These deviations show no correlation [for both positive and negative biases (Fig. 3C)] with the size of the local gap within our experimental error.

For boson-mediated pairing, variation of the pairing gap can be caused by changes in either the local boson energy or the local coupling between the boson and electrons (*2*). Such changes are reflected directly in the size and energy range of the strong-coupling features in the conductance ratio. Indeed, in metallic alloy systems (*39, 40*) where pairing is controlled by strong electron-phonon coupling, the magnitude of strong coupling features in the conductance ratio scales with the gap size [see supporting online material (SOM) text S1]. Because both the energy scale of the boson modes and the local electron-boson coupling do not correlate strongly with the magnitude of the local pairing gap in our samples, we conclude that the coupling to bosons in the range of 20 to 120 meV cannot be responsible for these inhomogeneous pairing gaps.

Although bosons may not be critical to pairing in $Bi_2Sr_2CaCu_2O_{8+\delta}$, the *R(r,V, T)* curves clearly show that these boson modes give a strong energy dependence to the gap function. Specifically, modification of Eq.2 using a complex energy dependent pairing gap $\Delta(r,\omega) = \Delta_R(r,\omega) + i\Delta_I(r,\omega)$, where $\Delta_R$ and $\Delta_I$ are the real and imaginary part of the gap function at energy $\omega$, can be used to capture the bosonic features in the conductance ratio at higher energies. Within such a model, we estimate that the interaction with bosonic excitations (in the range of 20 to 120 meV) results in a substantial imaginary component of the pairing interaction (about 25 meV in magnitude at $E$ = 40meV).

**Spatial Structure of Normal State Excitations & Inhomogeneous Pairing Interaction**

In search of the origin of the inhomogeneity in the pairing interaction in the cuprates, we focus on spectroscopic measurements of the electronic excitations in the normal state and their correlation with the inhomogeneity in the superconducting gaps. To reach the normal state, the temperature has to be high enough such that all the local pairing gaps have collapsed. For overdoped $Bi_2Sr_2CaCu_2O_{8+\delta}$ samples (hole doping x=0.24, $T_c$=62 K), less than 1% of the sample shows a gap at 90 K. In the intermediate temperature between $T_c$ and 90K, these samples show a mixture of ungapped and partially gapped spectra, as previously reported (*24*). Above 90K, the tunneling spectra, (Fig. 4A), are gapless at all locations on the sample but show asymmetric behavior for electron and hole tunneling. Careful examination of these spectra, over a wide range of energies, shows that electronic excitations in the sample are still spatially inhomogeneous at temperatures well above when pairs first form in the sample. The spatial inhomogeneity of the normal state's electronic excitations can be measured using conductance *(dI/dV)* maps at various voltages, which show variations on the length scale of order 20 Å (Fig. 4B and D to I). The magnitude of the variations is strongest for the conductance map obtained at the Fermi level (Fig. 4B), but such variations persist up to a few hundred meV.

The spatial variations of the normal state conductance can be compared with the low-temperature variations of the gap by using our lattice tracking technique. Shown in Fig. 4C is a gap map measured at 50 K over the exact same area of the sample as Fig. 4B (*41*). We can see a marked similarity between this gap map and the conductance map at the Fermi energy (Fig. 4B)—regions with a lower normal state conductance at the Fermi level nucleate superconducting gaps at higher temperatures resulting in larger low-temperature gaps. Quantifying these correlations in Fig. 5A, we show that the normal state conductance map and the low temperature gap map are strongly anti-correlated (-0.75). Further, both these maps have very similar auto-correlation lengths indicating that the spatial variation of the normal state is intimately linked to that of the low temperature gap. These measurements show that the variation of the superconducting state in $Bi_2Sr_2CaCu_2O_{8+\delta}$ samples, which for typical superconductors is characterized by a temperature-dependent superconducting coherence length, appears to be for the most part determined by the spatial variation of the normal state at high temperatures.

Because our measurements show that the spatial variation of the normal state conductance and the low temperature pairing gap maps are intimately connected, we can associate an average normal state spectrum with a low temperature gap value. We thus average together normal state spectra of regions of the sample that show identical low temperature gaps, and plot these average spectra in Fig. 5B. Systematic differences in the normal state spectra foreshadow the eventual variation of the gap in the superconducting state. In particular, the systematic shift of a "hump" in the normal state tunneling spectra at negative bias, in the range of -150 to -300 meV, (Fig. 5C) as well as the value of tunneling conductance at the Fermi energy (Fig. 5B, inset) track the size of the superconducting gap observed at low temperatures. While both the tunneling matrix element and the density of states of the tip influence the shape of the tunneling conductance in the normal state, the features of the normal state and the correlation (Fig. 5) have been observed in measurements using several different micro-tips (*42*).

It is important to compare our measurements of the normal state with those obtained from other spectroscopic techniques. In both angle-resolved photoemission and optical spectroscopy, strong renormalization of the single-particle excitations has been observed over an energy range of ~200 to 400 meV below the Fermi energy in Bi$_2$Sr$_2$CaCu$_2$O$_{8+\delta}$ samples (*17-20, 38, 43-45*). Such effects have been interpreted as either due to the coupling of electrons with a spectrum of bosonic excitations (such as spin fluctuations) or as a consequence of the energy band structure (such as the bilayer splitting) of this compound (*19, 46*). Although strong electron-boson coupling can be expected to modify the shape of the normal state spectrum, it is difficult to associate the normal state features that we measure with coupling to bosonic excitations due to their strong electron-hole asymmetry. Although we cannot rule out bosonic excitations as the origin of these features, candidate bosons would have to couple very asymmetrically to the tunneling of electrons and holes. Assigning these features to effects calculated from a simple non-interacting band structure can also be questioned given both the strong spatial variation at the atomic scale of the normal state spectra reported here, and the strong renormalization of single-particle states at similar energies in other spectroscopic studies of the normal state. Instead, these features might be the excitations of a doped Mott insulator where the electron and hole excitations are naturally asymmetric as a consequence of the strong Coulomb interaction (*47-49*). Some recent calculations indeed produce a "hump" in hole-like excitations that correlate with the strength of pairing (*47, 49*). Although there is no clear consensus on the right model for these excitations, our experiments show that the spectroscopic features of this state are indeed the origin of the nanoscale variation of the pairing strength in the superconducting state. Further experiments in samples at different hole doping levels at higher temperatures will be required to provide the detailed evolution of these atomic scale spectroscopic features of the normal state across the phase diagram.

**Concluding Remarks**

From a broader perspective, we have used the spatial variation of the pairing gaps, which gives rise to a range of pairing temperatures in nanoscale regions of our samples above $T_c$, as a diagnostic tool to find clues to the underlying mechanism of superconductivity. Temperature-dependent lattice tracking spectroscopy has allowed us to demonstrate that electron-boson coupling in the 20 to 120 meV range does not cause the variation of pairing gaps and onset temperatures in our samples. In contrast, we find that the high-energy (up to ~ 400 mV) hole-like excitations of the normal state are a direct predictor of strength of pairing and its spatial variation. The anticorrelation between the normal state conductance at the Fermi level and local strength of pairing also runs contrary to a BCS-like pairing mechanism where the coupling to bosons is proportional to the density of states at the Fermi energy (*2*).

Finally, we address the question of the underlying cause of variations of the normal state excitations in Bi$_2$Sr$_2$CaCu$_2$O$_{8+\delta}$ samples. Our analysis finds that both structural and electronic features of the samples contribute to such variations. We find that there are small correlations (about 10%) of the normal state conductance maps with the structural supermodulation along the b-axis in these samples. Similarly, we find that maps of electronic resonances around -900 meV previously probed in similar samples with STM (*50*) are correlated with the normal state conductance maps (about 30%) (SOM). Our measurements show that structural and chemical inhomogeneity affects both the excitations of the normal state and the superconducting gap. As is common to several correlated systems, many structural and electronic features can influence the onset and strength of collective phenomena (*51*). Our ability to correlate nanoscale excitation spectra between two distinct electronic states at the same atomic site provides the ability to study correlated phenomena in compounds with heterogenous chemical and structural properties.

52. Our studies were inspired by discussions with D. J. Scalapino and P. W. Anderson. We also thank R. Melko and D. J. Scalapino for help with the analysis of the low-energy electron-boson coupling. The work at Princeton is supported by DOE under contract DE-FG02-07ER46419 and NSF through the Princeton Center for Complex Materials. Y.A. was supported by KAKENHI 19674002. The work in BNL is supported by DOE under contract DE-AC02-98CH10886.

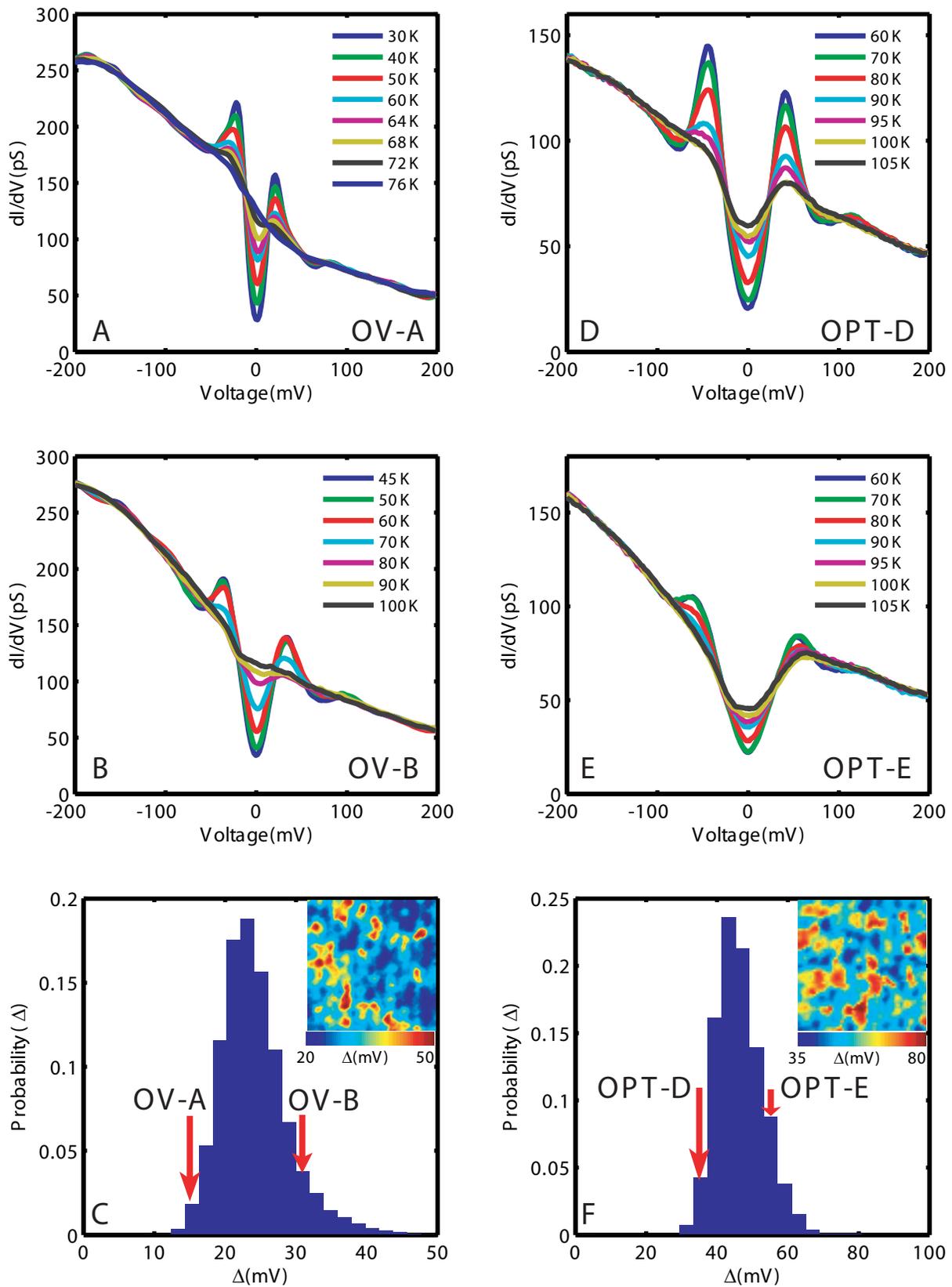

**Fig.1.** (**A** and **B**) Spectra taken at two different atomic locations on an overdoped $Bi_2Sr_2CaCu_2O_{8+\delta}$ sample ($T_C$=68 K, OV68) at various temperatures. The gaps in the spectra close at different temperatures, leading to a temperature independent background conductance at high temperature. (**C**) Histogram of pairing gap values measured in the OV68 sample. (Inset) A typical pairing gap map (300 Å) obtained for an OV68 sample at 30 K. (**D** and **E**) Spectra taken at two different atomic locations on an optimally doped sample ($T_C$=93 K, OPT) at various temperatures. The background continues to be temperature dependent well above $T_C$. (**F**) Histogram of gap values observed in the OPT sample. (Inset) typical gap map (300 Å) obtained for an OPT sample at 40 K.

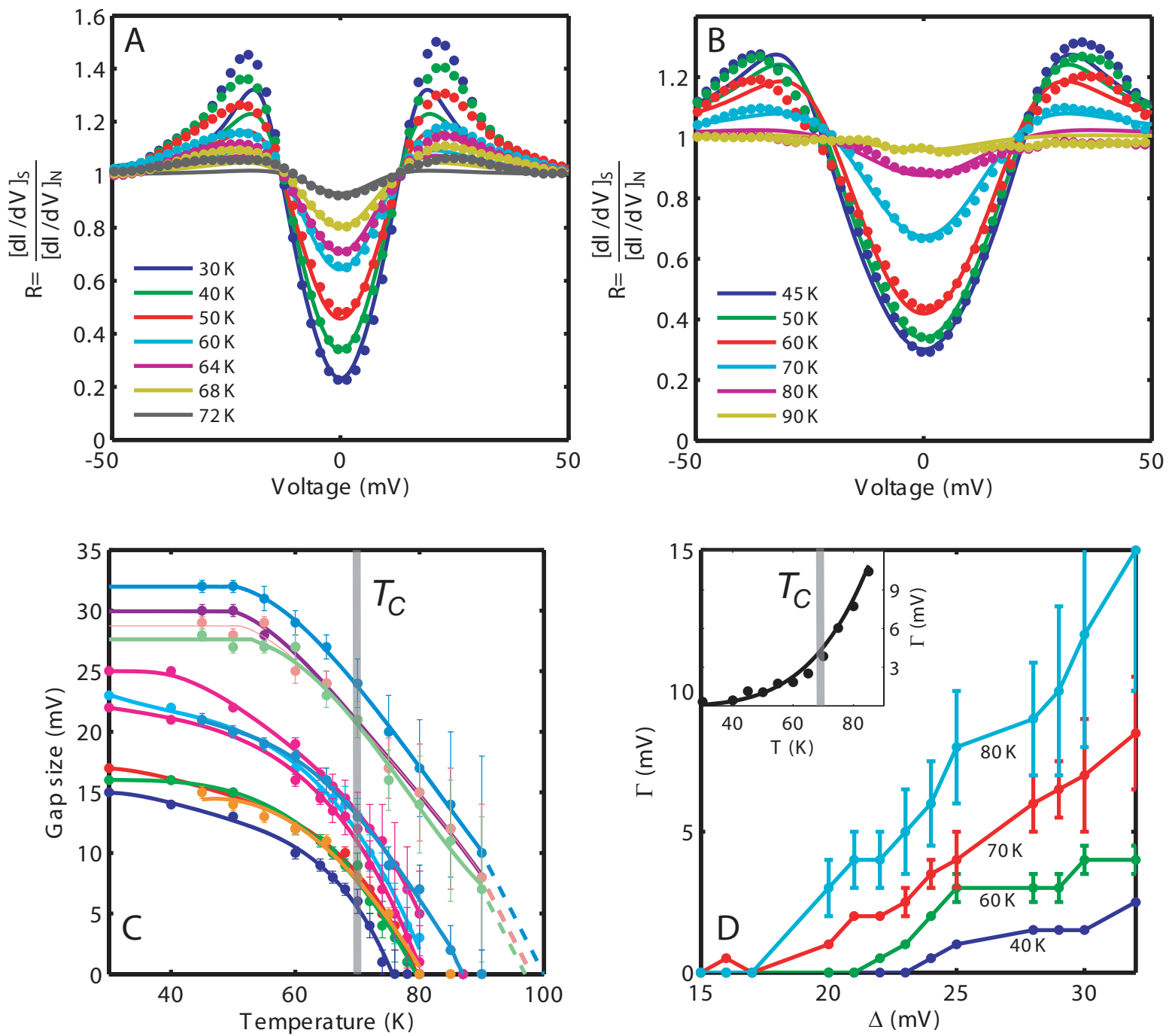

**Fig. 2.** (**A** and **B**) (data points) The conductance ratio R=(dI/dV(V,T))/(dI/dV(V,T>>$T_C$)) obtained by dividing the raw spectra in **Fig 1A** and **B** by the high temperature background spectrum. (lines) fits to the conductance ratio using a thermally broadened d-wave BCS model. In general, the fits work well at low V where the slope of the conductance ratio is inversely proportional to the gap. (**C**) Extracted values of the pairing gap for several different locations plotted as a function of temperature. The resistive $T_C$ is indicated by the grey line. (**D**) Extracted lifetime broadening at different temperatures plotted as a function of the corresponding low temperature gap. (Inset) the average lifetime broadening as a function of temperature.

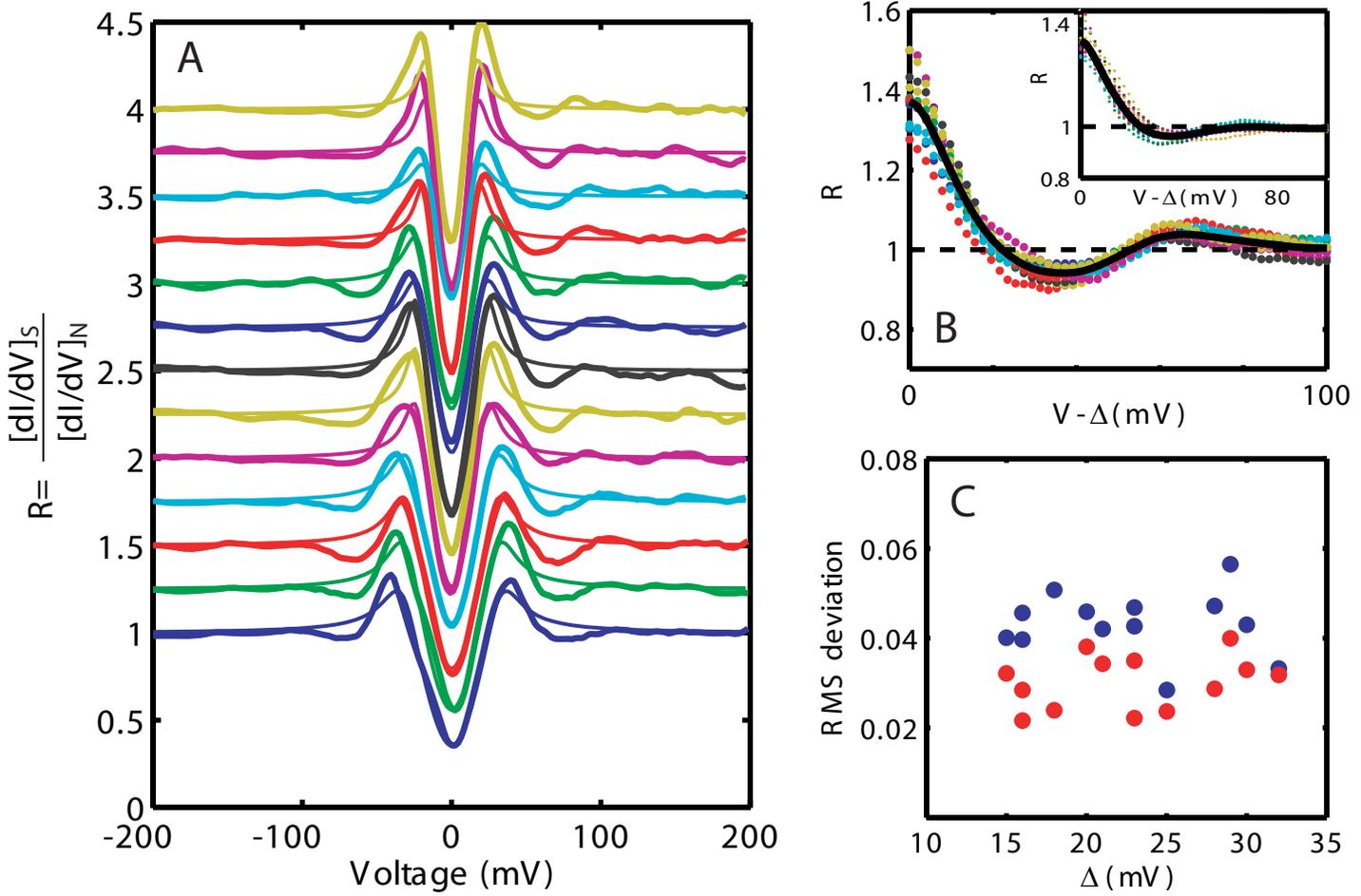

**Fig. 3.** (**A**) The low temperature (T=30 K) conductance ratio plotted for several different gaps. The conductance ratios deviate systematically from the d-wave model (equation 1, thin lines) and go below unity over a range of voltages (50-80 mV), indicating the strong coupling to bosonic modes. (**B**) The positive bias conductance ratios referenced to the local gap at different locations showing that the magnitude of the dip-hump feature is similar at all locations. The line is the average of all the locations. (Inset) Gap referenced conductance ratios for negative bias. (**C**) The RMS deviation of the conductance ratios from the d-wave model for positive (blue) and negative (red) bias over the energy range 20-120 mV. No correlation is seen between the magnitude of the deviations and the size of the gap.

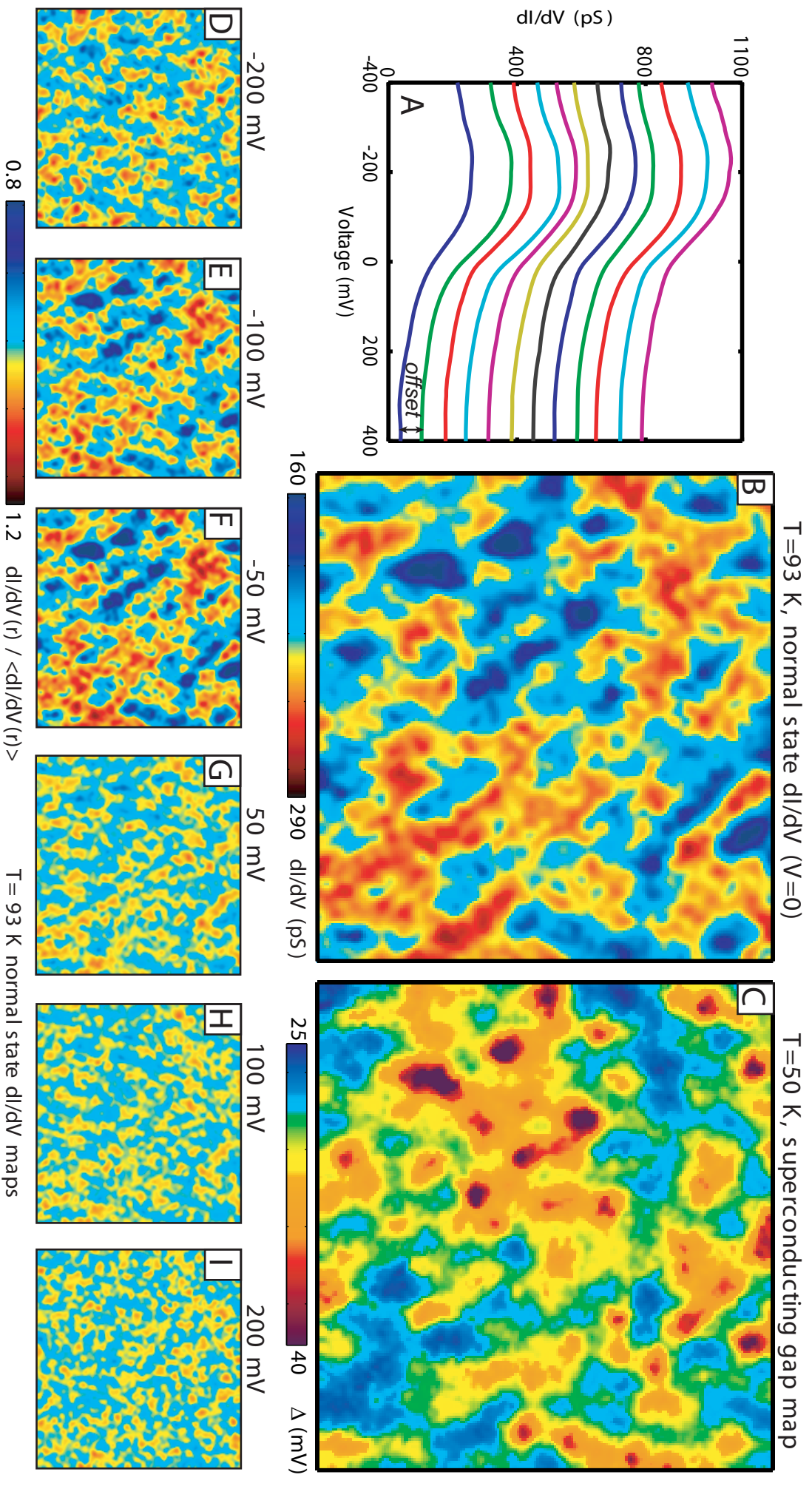

**Fig. 4.** (**A**) Spectra obtained at evenly-spaced locations along a 250 Å line in the normal state (T=93 K) of an OV62 sample. Less than 1% of the sample shows a remnant of a gap at this temperature. (**B**) Differential conductance map at the Fermi energy obtained at 93 K (**C**) Low temperature (50 K) gap map obtained on the same area as (**B**). (**D-I**) Spatial maps of the conductance at different energies obtained in the normal state. The junction is stabilized at +1V and 40 pA where there is minimum topographic disorder. While conductance maps at high energies show mostly structural features (b-axis supermodulation), the low energy spectra are inhomogeneous on the ~15 Å length scale.

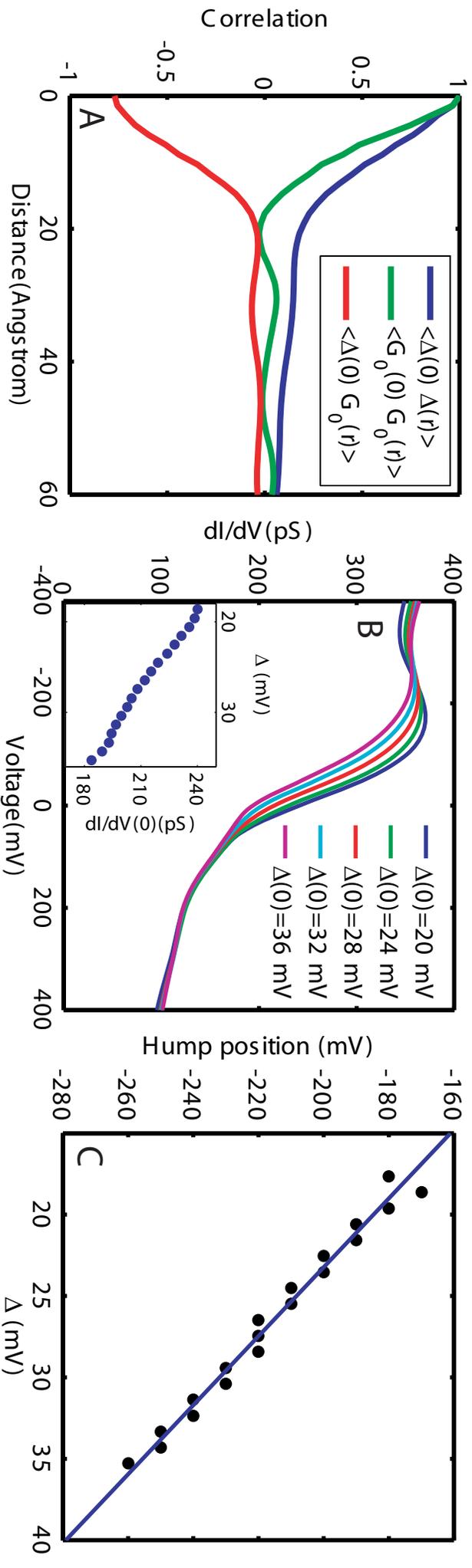

**Fig. 5. (A)** Angle-averaged autocorrelation of the low temperature gap map in Fig. 4C (blue line), autocorrelation of the conductance map in Fig. 4B (green line) and cross correlation between the two images (red line). We see that all the correlation lengths are similar (~ 15 Å) and that there is a strong anticorrelation between the normal state Fermi level conductance and the low temperature gap map. **(B)** Average normal state (T=93 K) spectra measured in different regions that show distinct low temperature superconducting gaps $\Delta_0$. Systematic changes are seen in the shape and position of the "hump" feature seen for the hole-like excitations (Inset) Differential conductance of the normal state at the Fermi energy as a function of $\Delta_0$. **(C)** The energy corresponding to the "hump" feature in the spectra as a function of $\Delta_0$.

# SUPPORTING ONLINE MATERIAL

## 1. Relationship between the conductance ratio and the boson coupling constant

In the Eliashberg theory of superconductivity, the pairing gap Δ is given by *(S1)*

$$\Delta = \hbar\omega_c \exp\left(\frac{\lambda - \mu^*}{1 + \lambda}\right)$$

Here $\hbar\omega_c$ is a cutoff frequency related to the boson density of states, $\lambda$ is the effective coupling constant and $\mu^*$ is the Coulomb pseudopotential. In order to increase the size of the gap and Tc, previous experiments on metallic alloys have focused on either increasing the energy range over which the bosons are coupled to the electrons ($\hbar\omega_c$) or alternatively, increasing the coupling to the bosons $\lambda$.

Strong coupling ($\lambda \sim 1$) to a bosonic mode at ω results in structure in the tunneling density of states at an energy Δ+ω. A change in the energy of the boson shifts the energy scale at which the strong coupling features are observed relative to the gap. On the other hand, increasing the coupling constant does not change the energy at which the features are observed but instead magnifies the size of the strong coupling features seen in the tunneling conductance. In the Lead-Thallium-Bismuth alloys, tuning the percentage of thallium and bismuth in lead does not lead to a significant change in the phonon energies *(S2)*. However, the coupling constant increases monotonically on changing the alloy composition from $Pb_{0.6}Tl_{0.4}$ through Pb to $Pb_{0.65}Bi_{0.35}$. Such changes in the coupling constant result in a corresponding increase in the magnitude of the strong-coupling features seen in the tunneling conductance (Fig. S1A, from *(S2)*). It is seen that (Fig. 2B) the RMS value of the deviations from the BCS density of states (over the energy range 0-10 mV) scales with the size of the gap (Fig. S1B). The absence of such scaling for the low-energy boson modes in $Bi_2Sr_2CaCu_2O_{8+\delta}$ indicates that the coupling constant is not increasing proportional to the gap siz

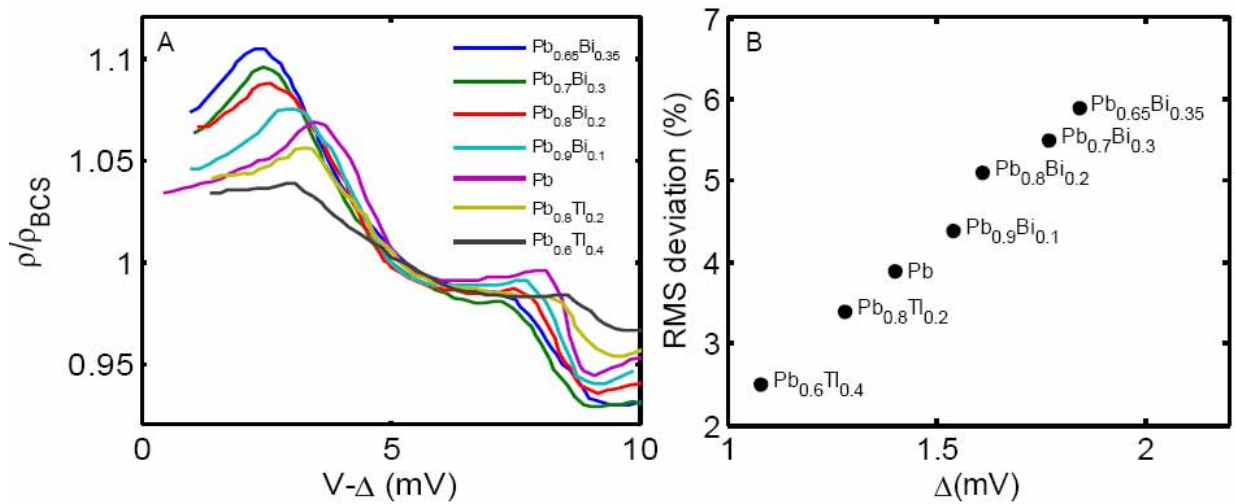

Figure S1: A. Strong-coupling deviations from the BCS theory for a series of lead-based alloy superconductors (from *(S1)*). While the phonon energy does not change significantly across the different materials, the coupling to the phonon increases. This increase causes a corresponding increase in the size of the deviations. B. RMS value of the deviations in the energy range 0-10 mV for the different samples in A. As the size of the deviations increases, the value of the superconducting gap also increases.

## 2. Correlation between the normal state, b-axis supermodulation and -900 mV resonances

Previous measurements in the superconducting state of $Bi_2Sr_2CaCu_2O_{8+}$ *(S3, S4)* have shown that the tunneling gap is correlated to inhomogeneity in the structure and electronic properties. In particular, it has been shown that the b-axis supermodulation running through the crystal affects the gap by about 10% *(S3)*. It has also been shown that the location of high-energy resonances (previously identified with oxygen dopants) anticorrelates by ~ 30% with the gap magnitude *(S4)*. Our measurements show that these structural and electronic features also affect the normal state spectrum. Shown in Fig. S2A is a gap map of an overdoped $Bi_2Sr_2CaCu_2O_{8+}$ sample (Tc=60 K) taken at 50 K. Shown in Fig. S2B is the conductance at the Fermi level measured in the normal state (T=95 K). Fig. S2C shows the topograph taken at high positive bias (800 mV) in the same area of the sample showing the b-axis supermodulation. To show the correlation between these quantities, we average these maps along the a-axis. The resultant variation in the gap, normal state conductance as well as the topographic height is shown in Fig. S2E, showing clearly that both the gap as well as the normal state conductance are affected by the b-axis supermodulation.

Shown in Fig. S2D is a map of the conductance at -900 meV taken in the normal state showing a number of resonances. These resonances affect both the gap map as well as the normal state conductance as shown in the correlation functions plotted in Fig. S2F. While the resonances have a ~ 30% effect on gap and the normal state, the normal state anticorrelates much more strongly with the gap map as discussed in the text. Our conclusion from such studies is that while a variety of factors can influence the low temperature gap, changes in the gap are foreshadowed by similar changes in the normal state excitation spectrum of $Bi_2Sr_2CaCu_2O_{8+}$ .

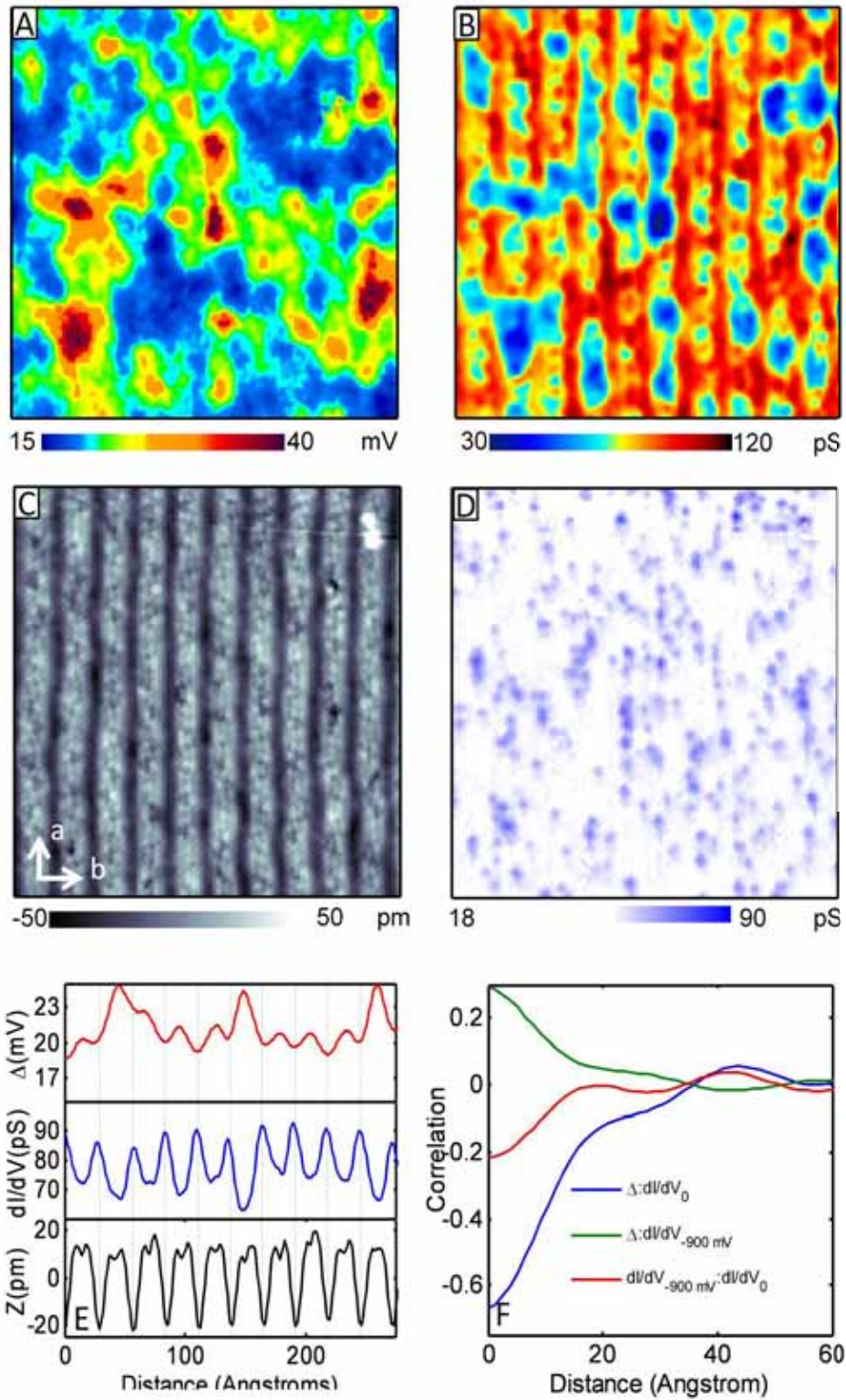

**Figure S2: A**. Gap map of an overdoped (Tc=60 K) sample taken at 50 K. **B**. Conductance at the Fermi energy (dI/dV$_0$) taken in the normal state (T=95 K). **C.** Topograph taken at +800 mV sample bias showing the b-axis supermodulation. **D**. Conductance at -900 mV showing resonances scattered in space. **A-D** are acquired in the same area of the sample. **E.** The variation in the gap map, normal state conductance map and topography after averaging along the a-axis. The gap map shows an ~ 10% change from peak to trough of the supermodulation. **F.** Correlation between the -900 mV resonances and the gap map (green) and between the -900 mV resonances and dI/dV$_0$ (red). Also shown for comparison (blue) is the correlation between dI/dV$_0$ and the gap map.